\documentclass[12pt]{article}
\begin{document}
\begin{large}
%
\catcode`@=11
\newcount\@tempcntc
\def\@citex[#1]#2{\if@filesw\immediate\write\@auxout{\string\citation{#2}}\fi
  \@tempcnta\z@\@tempcntb\m@ne\def\@citea{}\@cite{\@for\@citeb:=#2\do
    {\@ifundefined
       {b@\@citeb}{\@citeo\@tempcntb\m@ne\@citea\def\@citea{,}{\bf ?}\@warning
       {Citation `\@citeb' on page \thepage \space undefined}}%
    {\setbox\z@\hbox{\global\@tempcntc0\csname b@\@citeb\endcsname\relax}%
     \ifnum\@tempcntc=\z@ \@citeo\@tempcntb\m@ne
       \@citea\def\@citea{,}\hbox{\csname b@\@citeb\endcsname}%
     \else
      \advance\@tempcntb\@ne
      \ifnum\@tempcntb=\@tempcntc
      \else\advance\@tempcntb\m@ne\@citeo
      \@tempcnta\@tempcntc\@tempcntb\@tempcntc\fi\fi}}\@citeo}{#1}}
\def\@citeo{\ifnum\@tempcnta>\@tempcntb\else\@citea\def\@citea{,}%
  \ifnum\@tempcnta=\@tempcntb\the\@tempcnta\else
   {\advance\@tempcnta\@ne\ifnum\@tempcnta=\@tempcntb \else \def\@citea{--}\fi
    \advance\@tempcnta\m@ne\the\@tempcnta\@citea\the\@tempcntb}\fi\fi}
\catcode`@=12
\linespread{1.3}
\title{A solution of the spacetime singularity problem in relativistic
cosmology by using an additional variable\footnote{revied version}}
\author{{\sc Jae-Hyung ~Myung}\\
II.Institut f\"ur Theoretische Physik, Universit\"at Hamburg\\
D-22761 Hamburg, Germany}
\date{01/March/1996}
\maketitle
\begin{abstract}
The spacetime singularity in relativistic cosmology is cancelled by
using an additional variable. That is, the singularity-free models for
an expanding universe are obtained from general relativity.
\end{abstract}
%
\newpage
\section{Introduction}
Many physicists believe that the Friedmann models~\cite{Friedmann:1922,Friedmann:1924,Einsdesitt} 
describe the expanding phenomena of the space of the universe, and they accept
the models as a standard cosmological model. However, the models have many problems. The scale 
factor R in the models has no exact meaning and is an unobservable quantity, that is, R is not 
an astrophysical quantity but only a mathematical one. Thus the models do not agree with
Hubble's observations~\cite{Hubble:1929,Hubble:1931} completely. The Hubble's law ($V=H_{0}
D$, Fig.1) states that the distance between two neighbouring galaxies increases with time. If the 
universe is homogeneous and isotropic, the increasing distance means that the radius or the diameter 
of the universe is increasing. Therefore, the scale of the radius of the universe depends on the 
distance between two neighbouring galaxies. However, the Friedmann models do not represent this fact.\\
The energy tensors ($T_{ik}$) in the Friedmann models only have a contribution from the mass density 
of the galaxies. Therefore, the models should only describe the expansion of the present universe.
However, the models are also used to expansion of the early universe.\\
An expansion age of the universe estimated by the Friedmann model does not agree with the experimental 
data~\cite{Narlikar:1983}, that is, the age of the Friedmann universe is smaller than that of the 
components (stars or galaxies) of the real universe. Recent measurements using the Hubble
Space Telescope~\cite{Freedman:1994} and the Canada-France-Hawaii Telescope~\cite{Pierce:1994}
have given the Hubble constant $H_0$=80 $Kms^{-1}Mpc^{-1}$ and 87 $Kms^{-1}Mpc^{-1}$, respectively. 
According to the Friedmann model these values give an expansion age of $8\times10^{9}yrs$ and
$7\times10^{9}yrs$, respectively. Theses ages lie below the age of globular clusters ($10-18\times10^{9} yrs$) 
obtained from stellar evolution theory~\cite{Peebles:1993}.\\
The most important problem in the Friedmann models is the spacetime singularity. First, it is very 
difficult to imagine a universe that has a zero radius at the time zero. That is, with a gravitation theory 
we cannot explain the creation of the space of the universe, since the theory describes the interaction of 
matter. We must define the concept of the universe. The universe is a very large group of $10^{11}$
galaxies in an infinite space, and the galaxies in the components of the universe, i.e. local group, 
cluster and supercluster, are connected through gravitation. The space of the universe is just a space 
filled by these galaxies. Therefore, the space of the universe is only a subspace of the infinite space 
and cannot have been created by the big bang or a physical process. However, the spacetime singularity 
confuses this concept of the universe. Thus a few astronomers~\cite{Bondi:1948,Hoyle:1948} proposed
the steady state theory for an expanding universe.\\
Second, general relativity cannot be applied to the early universe in particle form, since the particle is 
the source of neither the weak nor the strong field. However, the spacetime singularity compelled us to 
apply the theory to the early universe. Third, according to the big bang model the temperature of the 
universe was nearly infinite at the big bang epoch. However, the universe should have an upper limit in 
temperature. Fourth, the singularity gave rise to the horizon and the flatness problems in the initial 
phase of the universe. The inflationary universe~\cite{Dolgov:1990,Guth:1981,Linde:1984} is an 
alternative model for a solution of these problems. However, the model cannot fundamentally solve the 
singularity problem.\\
If a model for an expanding universe agrees completely with the Hubble's law, the model should have no 
singularity, since the Hubble's law has no singularity and is valid only for the expanding phenomena of
the present universe. If the singularity is due to general relativity~\cite{Hawking:1970}, the 
theory is not a complete theory, since a gravitation theory cannot describe the creation and the annihilation 
of the space of the universe. However, the singularity is due not to the incompleteness of
general relativity but to a careless investigation of the spacetime geometry. We can avoid the singularity 
if we study the geometry carefully.
\section{New geometrical investigation of the present universe}
The main purpose of relativistic cosmology is to describe how $10^{11}$ galaxies move relative to the centre 
of the universe. We assume that the galaxies are isotropically and homogeneously distributed in the space
of the universe~\cite{Milne:1934}, and we take the Weyl postulate~\cite{Weyl:1923} for
the simplicity of calculations. Under these assumptions, the line element in four-dimensional spacetime 
coordinates $x^{0},x^{1},x^{2}$ and $x^{3}$ has the following form
\begin{equation}
ds^{2}=\left(dx^{0}\right)^{2}-d\sigma^{2},
\end{equation}
where $d\sigma^{2} = \sum g_{ik}dx^{i}dx^{k}$ (i,k=1,2,3) and is the metric on one of the spherical 
hypersurfaces othogonal to the world line of galaxies.\\
We must express the metric $d\sigma^{2}$ in spherical coordinates. To to this, we consider a hypersphere of
 the radius R and embed it in the Cartesian coordinates. In terms of the Cartesian coordinates
$x_{1}, x_{2}, x_{3}$ and $x_{4}$, the equation of the surface of the four-dimensional hypersphere with constant 
positive curvature is given by
\begin{equation}
x_{1}^{2} + x_{2}^{2} + x_{3}^{2} + x_{4}^{2} = R^{2},
\end{equation}
where the radius R of the hypersphere is a constant in the Cartesian coordinates and corresponds to the 
radius of universe in the spacetime coordinates. If we choose the following coordinates \\
$x_1 = R \sinh \alpha \cos \theta$, $x_2 = R \sinh \alpha \sin \theta \cos \phi$,
$x_3 = R \sinh \alpha  \sin \theta  \sin \phi$, $x_4 = R \cosh \alpha$,\\
with $z = \sinh \alpha$ we get the metric
\begin{equation}\label{eq:eps}
d\sigma^{2} = R^{2}[\frac{dz^{2}}{1-z^{2}} + z^{2} (d\theta^{2} + \sin^{2}\theta
d\phi^{2})].
\end{equation}
$d\sigma^{2}$ is the metric on the surface of the hypersphere of the
radius R. In order to obtain a time-dependent metric $d\sigma^{2}$ in the
Cartesian coordinates, we must change the radius R. However, we cannot
directly insert R(t) instead of R in eq.~(\ref{eq:eps}), since R is a constant in the
Cartesian coordinates. We must distinguish a hypersphere in the
Cartesian coordinates from one in the spacetime coordinate. The
radius R should be a constant in the Cartesian coordinates. Therefore, we
must choose a hypersphere with the radius $\neq R$ and find it
in the real universe represented by the spacetime coordinates. That is,
we must find a time-dependent variable that is proportional to the
radius of the universe. According to the Hubble's law, the
variable is the distance r between two neighbouring galaxies. It can
be geometrically realized as follows.\\
Let us consider a cross section that contains the centre of the
hypersphere in the Cartesian coordinates. The cross section is a circle
of the radius R which corresponds to the cross section that contains
the centre of the universe in the spacetime coordinates. We must
consider the distribution of galaxies in the cross section. Since we
assumed an isotropic and homogeneous distribution of the galaxies, the
relation between R and r can be approximated in the cross section as
follows (see Fig.2)
\begin{equation}\label{eq:spd}
R = Ar + B,
\end{equation}
where A is the number of galaxies on an axis and B is the product of
A and D (diameter of galaxy). We must note that R is not a constant in
the spacetime coordinates.\\
We return to the Cartesian coordinates. Let us consider a cross
section of the radius r in the spacetime coordinates as in Figure 2.
The cross section of the radius r corresponds to that of the hypersphere
of the radius r in the Cartesian coordinates. From now on we consider a
hypersphere of the radius r instead of one of the radius R. We embed
the hypersphere of the radius r in the Cartesian coordinates.
Inserting r(t) instead of R in eq.~(\ref{eq:eps}), we obtain
\begin{equation}
d\sigma^{2} = r^{2}(t)[\frac{dz^{2}}{1-z^{2}} + z^{2} (d\theta^{2} + \sin^{2}\theta
d\phi^{2})].
\end{equation}
$d\sigma^{2}$ is the time-dependent metric on the surface of the
hypersphere of the radius r. r depends only on t and corresponds to the
distance between two neighbouring galaxies in the spacetime coordinates.
Since there is no relation between r and R in the Cartesian coordinates,
R is a constant. With the curvature parameter k we obtain a
time-dependent line element
\begin{equation}
ds^{2} = c^{2} dt^{2} - r^{2}(t)[\frac{dz^{2}}{1-kz^{2}} + z^{2} (d\theta^2 + \sin^{2}
\theta d\phi^{2})].
\end{equation}
With the line element we can solve the field equations of Hilbert~\cite{Hilbert:1915}
\begin{equation}
R_{ik}-\frac{1}{2}g_{ik}R = -kT_{ik}.
\end{equation}
Setting the spacetime coordinates ($x^{0},x^{1},x^{2},x^{3}$) and
the Cartesian coordinates ($ct,z,\theta,\phi$)as follows
\begin{equation}
x^{0}=ct,\quad x^{1}=z,\quad x^{2}=\theta,\quad x^{3}=\phi,
\end{equation}
 we get the nontrivial equations from the equations with
 $\dot{r} = dr/dt$
\begin{equation}\label{eq:cdu}
\frac{2\ddot{r}}{r} + \frac{\dot{r}^{2} + kc^{2}}{r^{2}}= \frac{8 \pi \gamma
T^{k}_{k}}{c^{2}}\qquad (k=1,2,3),
\end{equation}
\begin{equation}\label{eq:fdp}
\frac{\dot{r}^{2} + kc^{2}}{r^{2}} = \frac{8 \pi \gamma T^{0}_{0}}{3c^{2}}.
\end{equation}
The energy tensors of matter have the following forms
\begin{equation}
T^{k}_{k} = -p, T^{0}_{0} = \epsilon.
\end{equation}
In the case of the system of galaxy behaving like dust the energy
tensors (11) have specific forms
\begin{equation}
p = 0, \epsilon = \rho c^{2},
\end{equation}
where $\rho$ only has the contribution from the mass density of the
galaxies which is given by
\begin{equation}
\rho = \rho_{0} r^{3}_{0}/r^{3}.
\end{equation}
$\rho_{0}$ and $r_{0}$ are present values of $\rho$ and r. With the energy
tensors eq.~(\ref{eq:cdu}) and eq.~(\ref{eq:fdp}) become
\begin{equation}\label{eq:jfk}
\frac{2\ddot{r}}{r} + \frac{\dot{r}^{2}+kc^{2}}{r^{2}}= 0,
\end{equation}
\begin{equation}\label{eq:jhm}
\frac{\dot{r}^{2} + kc^{2}}{r^{2}} = \frac{8 \pi \gamma \rho_{0} r_{0}^{3}}{3r^{3}}.
\end{equation}
In the case of positive curvature (k=+1) eq.~(\ref{eq:jhm}) can be rewritten
\begin{equation}\label{eq:csu}
\dot{r}^{2} = c^{2}(\frac{\beta}{r}-1),
\end{equation}
with $\beta$ given by
\begin{equation}
\beta = \frac{2q_{0}}{\left(2q_{0}-1\right)^{3/2}}\frac{c}{H_{0}}.
\end{equation}
$H_{0}$ is the Hubble constant and $q_{0}$ is the deceleration parameter.
eq.~(\ref{eq:csu}) has the solution
\begin{equation}
r=\beta \sin^{2}\frac{\theta}{2}.
\end{equation}
This is the distance between two neighbouring galaxies. The distance
increases with time. This is the reason why the universe expands. The
radius R of the universe is obtained by inserting the solution in
eq.~(\ref{eq:spd})
\begin{equation}
R = A\beta \sin^{2}(\frac{\theta}{2}) +B.
\end{equation}
The radius oscillates between minimum value B and maximum $A\beta+B$,
that is, the galaxies go first away from the centre of the universe and
then move for the centre (Fig.3a).\\
In the cases of k=0 and k=-1 eq.~(\ref{eq:jhm}) can be solved. In these cases the
radius of the universe is illustrated in Fig.3b and Fig.3c. The universe
expands forever. The spacetime singularity exists no more in the radius
of the universe.
\section{Discussion}
We solved the spacetime singularity problem in relativistic cosmology
by using an additional variable r. The singularity is due not to the
incompleteness of general theory but to a careless investigation of the
spacetime geometry. The relativistic cosmology becomes complete with
the variable r.\\
Although the spacetime singularity was removed in the radius of the
universe, the singularity still exists in the distance between
two galaxies. The singularity means that the distance was zero after
formation of the galaxies and that they collide with each other again.
However, in the real universe the galaxies should have a minimum
distance after their formation, that is, the universe is expanding from
the minimum. Therefore, the singularity should be also removed. However,
it is impossible since general relativity is a relativistic theory of
gravitation in four dimensions.\\
The spacetime geometry does not explain why the distance 
between two galaxies increases or decreases, since general relativity
is a purely mathematical theory of gravitation. We need a
nonrelativistic and dynamical theory of gravitation for a more exact
description of the expanding and contracting phenomena of our universe~\cite{Myung}
 The physics of the early universe should be reconstructed in a
reasonable and scientifical manner.\\
\vspace{5mm}
Acknowledgement\\
This article will not be submitted to any journal.

\newpage
Figure captions\\
Fig.1 Hubble's law. (a) The recession velocity V of a galaxy is
      proportional to its distance D from the Milky Way $\bigoplus$.
      (b) Distribution of galaxies at a later time; the distance
      between two neighbouring galaxies increased.\\
Fig.2 Distribution of galaxies on an axis of the cross section that
      contains the centre of the universe in the spacetime
      coordinates.\\
Fig.3a Radius R of the universe as a function of cosmic time t for k=1.\\
Fig.3b Radius R of the universe as a function of cosmic time t for k=0.\\
Fig.3c Radius R of the universe as a function of cosmic time t for k=-1.
\end{large}
\end{document}